 \documentstyle[12pt]{article}

\let\a=\alpha \let\b=\beta  \let\d=\delta \let\e=\epsilon
  \let\q=\theta  \let\k=\kappa
\let\l=\lambda   \let\x=\xi \let\p=\pi 
\let\s=\sigma   \let\f=\phi  
\let\w=\omega      \let\G=\Gamma \let\D=\Delta

\let\la=\label  \let\re=\ref
  
 \def\bd{\begin{document}} \def\ed{\end{document}}
\def\ds{\documentstyle} \let\fr=\frac \let\bl=\bigl \let\br=\bigr
\let\Br=\Bigr \let\Bl=\Bigl
\let\bm=\bibitem
\let\na=\nabla
\let\pa=\partial \let\ov=\overline
\newcommand{\be}{\begin{equation}}
\newcommand{\ee}{\end{equation}}
\def\ba{\begin{array}}
\def\ea{\end{array}}
\newcommand{\ho}[1]{$\, ^{#1}$}
\newcommand{\hoch}[1]{$\, ^{#1}$}
\newcommand{\bea}{\begin{eqnarray}}
\newcommand{\eea}{\end{eqnarray}}
\newcommand{\ra}{\rightarrow}
\newcommand{\lra}{\longrightarrow}
\newcommand{\Lra}{\Leftrightarrow}
\newcommand{\ap}{\alpha^\prime}
\newcommand{\bp}{\beta^\prime}
\newcommand{\tr}{{\rm tr} }
\newcommand{\Tr}{{\rm Tr} }
\newcommand{\NP}{Nucl. Phys. }
\newcommand{\tamphys}{\it\hoch{2} Center for Theoretical Physics\\
Physics Department \\ Texas A \& M University
\\ College Station, Texas 77843}
\newcommand{\cern}{\it\hoch{1} Theory Division\\ CERN\\ CH--1211 Geneva 23}
\newcommand{\auth}{J. A. Dixon\hoch{2}, M. J. Duff{}\hoch{1,2}
and J. C. Plefka\hoch{2}\hoch{\ddagger}}

\begin{document}
\hfill{CERN-TH. 6614/92}

\hfill{CTP-TAMU- 60/92}

\vspace{24pt}

\begin{center}
{ \large {\bf PUTTING STRING/FIVEBRANE DUALITY TO THE TEST\hoch{\dagger}}}

\vspace{36pt}

\auth

\vspace{10pt}

{\cern}

\vspace{10pt}

{\tamphys}

\vspace{48pt}

\underline{ABSTRACT}

\end{center}

According to string/fivebrane duality, the Green-Schwarz factorization of the
$D=10$ spacetime anomaly polynomial $I_{12}$ into $X_4\, X_8$ means that just
as $X_4$ is the anomaly polynomial of the $d=2$ string worldsheet so $X_8$
should be the anomaly polynomial of the $d=6$ fivebrane worldvolume.
To test this idea we
perform a fivebrane calculation of $X_8$ and find perfect agreement
with the string one--loop result.

{\vfill\leftline{PACS numbers: 11.17+y, 11.10.Kk}\vfill
\leftline{CERN-TH. 6614/92}\leftline{August 1992}
\vskip 10pt
\footnoterule
{\footnotesize
 \hoch{\dagger} Research
                           supported in part by NSF Grant PHY-9106593
 \vskip -12pt
 \hoch{\ddagger} Supported by a Fulbright Scholarship}}
\baselineskip=24pt

\pagebreak

\setcounter{page}{1}

The dual formulations of $D=10$ supergravity, one with a 7--form field
strength\ho{1}
and the other with a 3--form field strength\ho{2} have long been something of
an
enigma from the point of view of superstrings. As field theories, each seems
equally as good. In particular, provided we couple them to $E_8 \times E_8$
or $SO(32)$ super Yang-Mills, then both are anomaly--free\ho{3}$^,$\ho{4}.
Since the 3--form
version corresponds to the field theory limit of the heterotic string, it was
natural to conjecture\ho{5} that the 7--form version corresponds to the field
theory limit of an extended object dual to the string: the ``heterotic
fivebrane''. Just as the 2--form potential $B_{MN}$ ($M=0,1,\ldots,9$) couples
to the $d=2$ string worldsheet via the term
\be
S_2 = {1 \over 2\p \ap} \int d^2 \x \, {1\over 2} \e^{ij} \pa_i x^M \pa_j x^N
 B_{MN}
 = {1 \over 2\p \ap} \int B_2
\la{1}
\ee
where $\x ^i$ ($i=1,2$) are the worldsheet coordinates and $(2\p\ap)^{-1}$
is the string tension, so the 6--form potential $B_{MNPQRS}$
($M=0,1,\ldots,9$) couples to the $d=6$ fivebrane worldvolume via the term
\bea
S_6 &=& {1 \over (2\p)^3 \bp} \int d^6 \x \, {1\over 6!} \e^{ijklmn}
 \pa_i x^M \pa_j x^N \pa_k x^P \pa_l x^Q \pa_m x^R \pa_n x^S
 B_{MNPQRS} \nonumber \\
   &=& {1 \over (2\p)^3 \bp} \int B_6
\la{2}
\eea
where $\x^i$ ($i=1,\ldots,6$) are the worldvolume coordinates and
$[(2\p)^3\bp]^{-1}$ is the fivebrane tension. Writing $H_3 = dB_2 + O(\ap)$
and $H_7 = dB_6 + O(\bp)$, the relation
\be
H_7 = e^{-\f}\, {\ast H_3}
\la{3}
\ee
where $\f$ is the $D=10$ dilaton and $\ast$ denotes the Hodge dual using the
canonical metric $g_{MN}({\rm can})$, ensures that the roles of field equations
and Bianchi identities of the 3--form version of supergravity are interchanged
in the 7--form version.
The main subject of this paper will be the $O(\ap)$ corrections to $H_3$ and
the $O(\bp)$ corrections to $H_7$, discussed below.
\par
  An existence proof for this heterotic fivebrane was provided by
Strominger\ho{6},
who showed that the heterotic fivebrane emerges as a soliton solution
of the 3--form version. He went on to suggest that the strong coupling regime
of the string should correspond to the weak coupling regime of the
fivebrane; an idea made quantitatively more precise in Ref.\ 7, where it was
shown that the $\s$--model metrics $g_{MN}({\rm string})$ and
$g_{MN}({\rm fivebrane})$ are related to the canonical metric by
$
e^{-\f /2}\, g_{MN}({\rm string}) = g_{MN}({\rm can}) = e^{\f /6}\, g_{MN}({\rm
 fivebrane})
$
and hence that
$
{\rm g}({\rm fivebrane}) = {\rm g}({\rm string})^{-1/3}
$
where ${\rm g}({\rm string})$ and ${\rm g}({\rm fivebrane})$ are the string
and fivebrane
loop expansion parameters. The same paper also established the Dirac
quantization rule
\be
2 \k^2 = n (2\p)^5 \ap \bp \qquad ,\, n={\rm integer}
\la{6}
\ee
where $\k^2$ is the $D=10$ gravitational constant.
Further evidence was provided by the complementary discovery that a heterotic
string emerges as a soliton solution of the 7--form version\ho{8}.
Both the string and fivebrane soliton solutions break half the supersymmetries,
 both saturate a Bogomol'nyi bound between the mass and the topological charge,
and both go over into the corresponding elementary string\ho{9} and elementary
fivebrane\ho{10} solutions at large distances. These elementary solutions are
the
extreme mass = charge limit of the black string and black fivebrane which
display event horizons and a singularity at the origin\ho{11}. However, they
are
mutually non--singular in the sense that the string is a non--singular solution
of the 7--form version and the fivebrane a non--singular solution of the
3--form version\ho{12}. Recent work on conformal field theories\ho{13} and
other
exact solutions\ho{14} are also all consistent with this string/fivebrane
duality
conjecture.
\par
Crucial to the solution of Ref.\ 8 was the observation that duality mixes up
string and fivebrane loops: what is a one loop effect for the string might be
a tree level effect for the fivebrane, and vice versa. At higher loop
orders, this leads to an infinite
number of non--renormalization theorems (including the vanishing of the
cosmological term) all of which are consistent with known string calculations
to higher orders both in $\ap$ (worldsheet loops) and ${\rm g}({\rm string})$
(spacetime loops)\ho{15}. It is this loop mixing which allows us to test
string/fivebrane duality, in spite of our ignorance of how to quantize the
fivebrane. If duality is correct, we should be able to reproduce string loop
effects from tree--level fivebranes !
\par
To see this, let us first consider the well--known $SO(32)$ Yang--Mills
Chern--Simons corrections to $H_3$. ($E_8 \times E_8$ requires a separate
treatment and will be discussed elsewhere).
In bosonic formulation, these are obtained\ho{16} by augmenting
the action $S_2$ of (1) by the WZW terms
\be
S_2{}^\prime = - n_2 \frac{6\p}{N_2} \int_{\pa M_3}\, \tr AK {} +
 n_2 \frac{2\p}{N_2} \int_{M_3}\, \tr K^3
\la{7}
\ee
where $A = A_M\, dx^M$, $K= g^{-1}dg$ and where the gauge fields $A_M=
A_M{}^a\,
T^a$ and the group elements $g$ are matrices in the fundamental representation
of $SO(32)$. Here $n_2 = {\rm
integer}$ is the level of the Kac--Moody algebra and $N_2$ is a normalization
constant given by the general formula\ho{17}
\be
N_{2n-2} = \Bl ( \frac{2\p}{i} \Br )^n
 \frac{(2n-1)!}{(n-1)!}
\la{8}
\ee
However, in $8k + 2$ dimensions for which $n=4k + 2$ we may define
Majorana--Weyl fermions and the corresponding WZW terms are to be divided by
2. This is the case for the heterotic string, so $N_2 = -48\p^2$.
This agrees with Witten\ho{18}. Let us define $F = dA + A^2$ and
\bea
I_{2n} &=& \Bl ( \frac{i}{2\p} \Bigr )^n \frac{1}{n!}\, \tr F^n  \nonumber \\
d \w_{2n-1} &=& I_{2n} \nonumber \\
\d \w_{2n-1} &=& d \w^1_{2n-2}
\la{9}
\eea
then the sum $S_2 + S_2{}^\prime$ is gauge invariant under
$\d A = d \l + [A,\l]$ and
$\d K = d \l + [K,\l]$
provided
\be
\d B_2 =  \frac{n_2}{2} \ap (2\p)^2 \w^1_2
\la{11}
\ee
and hence the gauge invariant field strength is given by\ho{19}
\bea
H_3 &=& dB_2 - \frac{n_2}{2} \ap (2\p)^2 \w_3 \nonumber \\
dH_3 &=& - \frac{n_2}{2} \ap (2\p)^2 I_4
\la{12}
\eea
This modification to the Bianchi identity is thus seen to be a classical string
effect (i.e. tree level in the $D=10$ string loop expansion).
\par
Next we turn to the Green--Schwarz anomaly cancellation mechanism\ho{3}
which is
a genuine quantum (string one loop) effect. The Majorana--Weyl gauginos of the
$D=10$ theory belong to the dimension 496 adjoint representation. The
non--abelian anomaly polynomial is given by the abelian anomaly in $d=12$,
which is given by $\frac{1}{2} I_{12}$ of (\re{9}) with the fundamental tr
replaced
by the adjoint Tr. The factor of $\frac{1}{2}$ arises because the fermions
are Majorana. As emphasized by Green and Schwarz\ho{3}, the miracle of $SO(32)$
is that since
$
\Tr F^6 = \Tr F^2\, \Tr F^4\,/48 - (\Tr F^2)^3\,/14,400$,
$I_{12}$ factorizes:
\be
\frac{1}{2} I_{12} = X_4 \, X_8
\la{15}
\ee
In fact, since $\Tr F^4 = 24 \,\tr F^4 + 3(\tr F^2)^2$ and
$\Tr F^2 = 30\, \tr F^2$,
we have the remarkable coincidence
\be
X_4 = \frac{1}{2}\, I_4 \qquad \qquad
X_8 =  I_8 \la{17}
\ee
The consistent anomaly
\be
G = \frac{1}{2} \cdot 2\p \int \Bl ( \frac{1}{3} \, \w^1_2 I_8
 + \frac{2}{3}\, \w^1_6 I_4 \Br )
\la{18}
\ee
is then cancelled by adding to the effective action
\be
\D \G_2 = - 2\p \int \Bl ( \frac{1}{n_2 \ap (2\p)^2} \, B_2\,
  I_8 + \frac{1}{3}\, \w_3 \, \w_7 \Br )
\la{19}
\ee
and recalling the transformation rule for $B_2$ given in (\re{11}). Now for
$B_2$ normalized as in (\re{1}), its kinetic term is
\be
\G_2 = - \frac{1}{2\k^2} \int \frac{1}{2} e^{-\f}\, H_3 \wedge \ast H_3
\la{20}
\ee
and hence the addition of (\re{19}) modifies the field equation to
\be
d(e^{-\f}\, \ast H_3) = \frac{2\k^2}{n_2 \ap (2\p)}\, I_8
\la{21}
\ee
\par
So far, all our considerations started with the string worldsheet. The acid
test for string/fivebrane duality is to reproduce (\re{21}) starting from the
fivebrane worldvolume. We begin by augmenting the action $S_6$ of (2) by
the WZW term\ho{20}
\be
S_6{}^\prime = \frac{70\p n_6}{N_6} \int_{\pa M_7} C_6\, -\, \frac{2\p
n_6}{N_6}
  \int_{M_7} \tr\,K^7
\la{27}
\ee
The explicit form for $C_6$ is given in Ref.\ 20. Here we need only note
that it
transforms as
\be
\d C_6 = 24 (2\p)^4 [\w^1_6(A,\l) - \w^1_6(K,\l)]
\la{28}
\ee
Here $n_6 = {\rm integer}$ is the level of the Mickelsson--Faddeev
algebra\ho{21}
and $N_6 = (2\p)^4 7!/3!$ from (\re{8}). The sum $S_6 + S_6{}^\prime$ is gauge
invariant provided
\be
\d B_6 = - n_6 \bp (2\p)^4 \w^1_6
\la{29}
\ee
and hence the gauge invariant field strength is given by\ho{19}
\bea
H_7 &=& dB_6 + n_6 \bp (2\p)^4 \w_7 \la{30a} \\
dH_7 &=& n_6 \bp (2\p)^4 I_8
\la{30b}
\eea
Using (3), this is identical to (\re{21}) provided
\be
2\k^2 = n_2 n_6 (2\p)^5 \ap\bp
\la{31}
\ee
But this is just the Dirac quantization rule (\re{6}) with $n= n_2 n_6$ !
\par
Having established that the classical fivebrane correctly reproduces the
quantum string result in the pure Yang--Mills sector, we now turn to the
gravitational and mixed anomalies. Now one must include the $D=10$ gravitino
contribution to $I_{12}$, but again it factorizes as in (\re{15}) where
now\ho{22}
\bea
X_4 &=& \frac{1}{2} \cdot \frac{1}{(2\p)^2} \Bl [ -\frac{1}{2}\,\tr F^2
+ \frac{1}{2}\,\tr R^2 \Br ] \nonumber \\
X_8 &=& \Bl (\frac{1}{2\p}\Br )^4 \Bl [ \frac{1}{24} \,\tr F^4 -
\frac{1}{192}\,
\tr F^2 \, \tr R^2 + \frac{1}{768}\,(\tr R^2)^2 + \frac{1}{192}\, \tr R^4
\Br ]
\la{32}
\eea
Since we have already obtained the correct overall normalization of $X_4$ and
$X_8$ it remains only to explain the relative coefficients. (We
set $n_2 = n_6 = 1$ from now on). This is most easily done by changing from
the bosonic WZW formalism to the fermionic one where $I_4$ and $I_8$ are
the anomaly polynomials arising via the index theorem from the chiral
fermions on the $d=2$ worldsheet and $d=6$ worldvolumes respectively.
We should mention that although we have the luxury of choosing a bosonic or
fermionic formulation on the $d=2$ worldsheet, there is probably no such
bose--fermi equivalence in $d=6$ and so ultimately one will have to choose
between the bosonic and fermionic formulations. This choice must await a
complete covariant $\k$--symmetric Green--Schwarz action for the heterotic
fivebrane which, to date, is still lacking\ho{23}. Fortunately, for the
present purposes of calculating
anomalies, either way will do. In the Yang--Mills case, this may be
seen explicitly via the
Euclidean identities for the index of the Dirac operator\ho{24}
\bea
({\rm ind}\, i D\!\!\!\! /)_{2n-2} &=& \int \, I_{2n} =
\Bl ( \frac{i}{2\p} \Br )^n
\frac{1}{n!} \int \,
\tr F^n \nonumber \\
&=& (-1)^{n-1} \Bl (\frac{i}{2\p} \Br )^n \frac{(n-1)!}{(2n-1)!}\, \int \,
\tr K^{2n-1}
\la{39}
\eea
which provide an independent check on the equivalence of the WZW and chiral
fermion calculations. For the gravitational anomalies, the WZW method is
unknown to us and we shall follow the fermionic approach where\ho{24}
\bea
I_4 &=& \frac{1}{(2\p)^2} \, \Bl [ \frac{i^2}{2}\,\tr F^2 + \frac{r}{48}\,
 \tr R^2 \Br ] \nonumber \\
I_8 &=& \frac{1}{(2\p)^4}\, \Bl [ \frac{i^4}{24}\,\tr F^4 + \frac{i^2}{96}\,
\tr F^2\,\tr R^2 + \frac{r}{4608}\,(\tr R^2)^2 + \frac{r}{5760}\,
 \tr R^4 \Br ]
\la{33}
\eea
Here the Yang--Mills trace is in whatever representation the fermions are
in and $r$ is its dimensionality and the Lorentz trace is in the vector
representation.
In the Green--Schwarz formalism the heterotic string is
described by superspace coordinates $(X^M , \q^\a)$ where the $\q$'s are in the
$16$ of $SO(1,9)$ and the $SO(32)$ quantum numbers are carried by
Majorana--Weyl fermions in the 32--dimensional fundamental representation.
Although the $\q$'s are worldsheet scalars, they are anticommuting and obey a
first order Lagrangian. However, because of $\k$ symmetry, only 8 of
the $\q$'s
are physical (equal to $10-2$, the number of transverse $X$'s). The net effect
of integrating out the $\q$'s is thus equivalent to 8 Majorana--Weyl fermions
but with the opposite chirality to the 32 gauge fermions\ho{25}. In summary we
calculate $X_4$ by dividing $I_4$ by two (because the fermions are Majorana),
taking the tr in the fundamental representation and setting $r = 32 -8 = 24$.
This agrees with (\re{32}). In the case of the fivebrane
we again have the same $(X^M , \q^\a)$ and we
again take the gauge fermions in the $32$ of $SO(32)$. Because we are in $d=6$,
 however, the fermions are no longer Majorana--Weyl (which exist only in
2 mod 8 dimensions) and so we
do not divide $I_8$ by 2. Moreover in $d=6$, the number of physical $\q$'s is
only 4 (equal to $10 - 6$ , the number of transverse $X$'s)
since $\k$--symmetry halves the number and (unlike $d=2$) going on--shell
halves it again. The net effect of integrating out the $\q$'s is thus
equivalent to only two $d=6$ fermions\ho{25} and hence $r = 32 -2
= 30$. Thus
\be
I_8 = \frac{1}{(2\p)^4}\, \Bl [ \frac{1}{24}\,\tr F^4 - \frac{1}{96}\,\tr F^2
 \, \tr R^2 + \frac{15}{2304}\,(\tr R^2)^2 + \frac{1}{192}\,\tr R^4 \Br ]
\la{34}
\ee
This is not quite in agreement with $X_8$ of (\re{32}) but the difference
is proportional to $\tr\, R^2\, X_4$ and hence the
discrepancy is
easily remedied. Up until now we have been assuming that the fivebrane $B_6$
which appears in (2) was identical to the $D=10$ supergravity $B_6$ which
satisfies (3). However, this may need to be modified when we include the
gravitational Chern--Simons corrections which are of higher order
in the low--energy expansion than those
of Yang--Mills.
The Green--Schwarz result is
$
dH_7({\rm supergravity}) = \bp (2\p)^4\, X_8$,
whereas our fivebrane calculations have lead us to
$
dH_7({\rm fivebrane}) = \bp (2\p)^4\, I_8
$.
So if we define
\be
B_6({\rm fivebrane}) = B_6({\rm supergravity}) - c\, \frac{\bp}{\ap}\, \tr R^2
\, B_2
\la{37}
\ee
then the gauge--invariant field strength $H_7({\rm supergravity})$
satisfies
\be
dH_7({\rm supergravity}) = \bp (2\p)^4\, (I_8 - \frac{c}{(2\p)^2}\,\tr\,R^2\,
 X_4)
\la{38}
\ee
on using (\re{12}).
Comparing $X_8$ of (\re{32}) with $I_8$ of (\re{34}) and using $X_4$ of
(\re{32}), we find perfect agreement with the Green--Schwarz result with the
choice $c=1/48$.
\par
In their discussion of anomalies, Green, Schwarz and Witten\ho{22} say
``The case where
something really new happens is that of $4k + 2$ dimensions and this is the
case
of interest to superstring theory since the worldsheet has dimension two and
the spacetime has dimension ten !''. The results of the present paper might
be summarized by adding ``$\ldots$ and the fivebrane worldvolume has dimension
six !''.
\bigskip
\par
\leftline{\bf Acknowledgements}
\par
Conversations with L. Alvarez--Gaum\'e, E. Bergshoeff, B. Campbell,
M. Grisaru, R. Stora
and P. Townsend
are gratefully acknowledged.
\pagebreak

\end{document}